\documentclass{llncs}

\usepackage{times}
\usepackage{psfig} 
\usepackage{latexsym} 
\usepackage{subfigure}
\usepackage{amssymb}
\usepackage{verbatim}
\usepackage{url}

\newcommand{\cv}{\ensuremath{\mathcal{V}}}
\newcommand{\ce}{\ensuremath{\mathcal{E}}}
\newcommand{\co}{\ensuremath{\mathcal{O}}}

\newcommand{\pre}{\ensuremath{\mathit{pre}}}
\newcommand{\post}{\ensuremath{\mathit{post}}}
\newcommand{\choice}[2]{\ensuremath{{#1}~||~{#2}}}
\newcommand{\ifthen}[2]{\ensuremath{{\bf if}\ {#1}\ {\bf then}\ {#2}}}
\newcommand{\ifthenelse}[3]{\ensuremath{{\bf if}\ {#1}\ {\bf then}\ {#2}\ {\bf else}\ {#3}}}
\newcommand{\while}[2]{\ensuremath{{\bf while}\ {#1}\ {\bf do}\ {#2}}}
\newcommand{\letin}[3]{\ensuremath{{\bf let}\ #1\ =\ #2\ {\bf in}\ #3}}
\newcommand{\myfigure}[4]{
\begin{figure}[t#1]
\framebox{\parbox{.98\textwidth}{#2\hfill}}
\caption{#3}
\label{#4}
\vspace{-30\in}
\end{figure}
}

\begin{document}

\title{Three-Tiered Specification of Micro-Architectures\thanks{This
research was supported, in part, by a Collaborative Research
Development grant from the Natural Sciences and Engineering Research
Council of Canada and Nortel, Nuns Island, Montreal, and by the Dutch
Research Organisation \emph{NWO}, in the project 612.014.006
``\emph{Generation of Program Transformation Systems}''.}  }

\author{Vasu Alagar$^{1,2}$
\and Ralf L{\"a}mmel$^{3,4}$}

\institute{${}^1$\ Concordia University, Montreal, Canada,
{\tt alagar@cs.concordia.ca}
\\
${}^2$\ Santa Clara University, Santa Clara, CA, USA,
{\tt valagar@cse.scu.edu}
\\
${}^3$\ CWI, Amsterdam, The Netherlands,
{\tt ralf@cwi.nl}
\\
${}^4$\ Free University, Amsterdam, The Netherlands,
{\tt ralf@cs.vu.nl}
}

\maketitle
\newlength{\lslwidth}
\setlength{\lslwidth}{0.2cm}
\newcommand{\key}[1]{{\small\bf #1}}
\newcommand{\lsllclindent}[1]{\hspace*{#1\lslwidth}}
\newcommand{\lslcommentline}[1]{\% #1}
\newcommand{\lslcomment}[1]{\hfill \% #1}
\newcommand{\lslskip}{\medskip}
\newenvironment{lsltrait}{
\renewcommand{\~}{\tilde{\;}}
\renewcommand{\^}{\hat{\;}}
\setlength{\parskip}{0in}
\setlength{\parindent}{0in}
\raggedright
}{}


\begin{abstract}
A three-tiered specification approach is developed to formally specify
collections of collaborating objects, say micro-architectures.  (i)
The structural properties to be maintained in the collaboration are
specified in the lowest tier. (ii) The behaviour of the object methods
in the classes is specified in the middle tier. (iii) The interaction
of the objects in the micro-architecture is specified in the third
tier. The specification approach is based on Larch and accompanying
notations and tools. The approach enables the unambiguous and complete
specification of reusable collections of collaborating objects.  The
layered, formal approach is compared to other approaches including the
mainstream UML approach.

\medskip

\end{abstract}

\section{Introduction}

\vspace{-20\in}

\paragraph{Class vs.\ micro-architecture vs.\ framework}

An object-oriented system is a collection of encapsulated objects that
collaborate among themselves to achieve specified tasks. The benefits
of systems designed using OO principles include the potential for
reuse, incremental extension and local modification. According to
\cite{LK95} and others, one can distinguish several levels of
reuse and adaptation. At the lowest level, one reuses or adapts
methods or \emph{classes}. This level is often inadequate for reuse,
and it does not appropriately scope adaptation activities since
methods and classes are not ``islands''. They cannot be reused and
adapted independently. At the highest level, one operates on entire
application \emph{frameworks}. This form does not account for
application development with reuse of building blocks, neither does it
restrict the scope of an adaptation. In fact, our target is the
intermediate level of \emph{micro-architectures}, that is, collections
of collaborating objects. We contend that this is the appropriate
level for \emph{reuse} and \emph{adaptation} in object-oriented design
and programming (OOD \& OOP).

\begin{figure}[ht]
\centerline{\psfig{figure=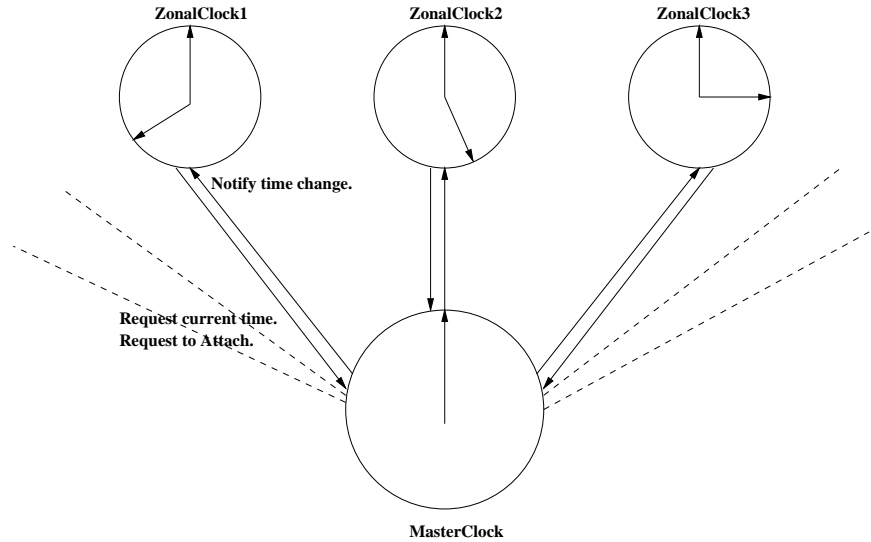,height=2.8in}}
\vspace{-20\in}
\caption{The micro-architecture \emph{WorldClock}}
\vspace{-40\in}
\label{F:ma}
\end{figure}

\paragraph{A running example}

Let us sketch an example of a micro-architecture. Our running
example is the \emph{WorldClock} micro-architecture dealing with
interacting \emph{MasterClock} and \emph{ZonalClock} objects as
illustrated in Figure~\ref{F:ma}. The \emph{MasterClock} object is
responsible for maintaining the Greenwich Meridian Time, while the
\emph{ZonalClock} objects display the time in their respective
zones. The \emph{MasterClock} object can exist independently of the
\emph{ZonalClock} objects, but each \emph{ZonalClock} object depends
on the \emph{MasterClock} object to maintain its zonal time. We thus
have a one-to-many relationship between master and zonal clocks. All
the objects together maintain an invariant that any zonal clock
displaying the time in its zone is consistent with the master clock's
time. The \emph{MasterClock} object notifies its associated
\emph{ZonalClock} objects whenever its time is updated. When a
\emph{ZonalClock} object is requested to update its zonal time,
it queries the \emph{MasterClock} object for the current time. If the
\emph{ZonalClock} object observes a time change then it updates itself
to make its state (i.e., the zonal time) consistent with the time
maintained by the \emph{MasterClock} object. Explicit polling by the
dependent \emph{ZonalClock} objects for new information is not
intended. The identities and the number of
\emph{ZonalClock} objects are not known a priori. Each \emph{ZonalClock}
object attaches itself to a \emph{MasterClock} object upon its
creation.

\vspace{-30\in}

\paragraph{Micro-architectures vs.\ design patterns}

Let us clarify our use of the term ``micro-architecture'' with regard
to the related term ``design pattern''~\cite{GHJV94}. Recall our
definition: micro-architectures correspond to collections of
collaborating objects. In fact, we consider micro-architectures to be
the building blocks of object-oriented applications, especially of
frameworks for domain-specific application development. A framework or
an application contains several micro-architectures. In the actual
code that embodies a framework, a certain class might contribute to
more than one micro-architecture. By contrast, design patterns
represent abstract, that is, application-independent
designs. Especially, the operational meaning of design patterns is
deliberately vague. One might say that design patterns are the most
abstract micro-architectures one can think of, and hence we might
consider micro-architectures as concrete instances of design patterns
within frameworks.  The typical micro-architecture in a framework is
more concrete than a design pattern because it will usually exhibit
some domain-specific behaviour. To give an example, the
\emph{WorldClock} micro-architecture is a concrete instance of the
\emph{Observer} pattern~\cite{GHJV94}. While the \emph{WorldClock}
micro-architecture deals with the notion of time based on
corresponding operations, the more abstract \emph{Observer} pattern
only involves an abstract notion of state.

\vspace{-30\in}

\paragraph{In need for a specification approach}

An informal explanation like the one given for the running example
above is certainly \emph{informative} for a developer who wants to
reuse or to adapt a micro-architecture. However, in order to
adequately deal with the complexity of design, improve productivity,
and maintain acceptable levels of software quality the developer
should also be provided with an \emph{unambiguous} description of
software components. Informal descriptions by themselves are grossly
insufficient to build future software architectures, such as those
being planned in strategic applications. The mainstream approach to
specify designs of such architectures is to use UML~\cite{UML13}. One
uses class diagrams to model the static structure of entities in the
design, and one describes the collaborations using object
collaboration diagrams. Specifications of object interfaces are given
in pseudo-code.  This approach does not provide a clear description of
object dependencies and the inter-object behaviours that are
maintained in the collaboration. Finally note that an UML-based
specification is typically not formal, although there is admittedly an
ongoing effort to provide a formal interpretation of certain UML
notations.

\vspace{-30\in}

\paragraph{Three-tiered specification of micro-architectures}

We contend that we need a \emph{complete} and \emph{formal} approach
to the specification of collections of collaborating objects. We also
want this approach to be \emph{simple} and to allow for a
\emph{seamless integration with UML} visual modelling facilities. A
corresponding specification approach is the prime contribution of the
present paper. Different aspects of micro-architectures are covered in
three tiers:
\vspace{-15\in}
\begin{description}
\item[(i)] the \emph{structural} properties to be maintained 
by the collaboration;
\item[(ii)] the \emph{roles} of the collaborating objects in a
black-box fashion;
\item[(iii)] the \emph{interactive} behaviour in terms of the operation
sequences and flow of control.
\end{description}\vspace{-15\in}
In our specification approach, we use a designated notation for each
tier. The fist two tiers are covered by the formal specification
language Larch~\cite{GH93}, namely Larch traits and the Larch/C++
notation. We chose Larch because Larch makes it possible to express
externally observable behaviour in an implementation independent
fashion which is crucial for black-box reuse. Larch/C++ provides
built-in syntactic and semantic support for specifying C++ class
interfaces. This allows us to link specification and programming.  As
for the third tier, we use a simple logic of actions in the style of
Lamport's Temporal Logic of Actions~\cite{Lamport94}.

\vspace{-30\in}

\paragraph{Structure of the paper}

The three tiers or our specification approach are presented in the
three sections \ref{S:I}--\ref{S:III} accordingly. We provide the
specification of the \emph{WorldClock} example throughout these
sections while the mainstream UML approach is considered as well. In
Section~\ref{S:related}, we report on related work regarding the
specification of design structures such as micro-architectures or
design patterns. In Section~\ref{S:conclusion}, we conclude the paper.

\vspace{-20\in}

\section{Lowest tier: Structure}
\label{S:I}

The lowest of the three tiers specifies the structural aspects of the
collaboration.  Besides specifying the \emph{data models} for the
objects in the collaboration, it also specifies the \emph{states of
interest} of each object, and the \emph{cardinality constraints} of
the relationships among the collaborating objects.  We use the Larch
Shared Language LSL \cite{GH93} for the specification of this
layer. The section is structured as follows. We first recall the
diagrammatic approach of UML corresponding to this layer. Then, we
provide a brief description of LSL. Finally, we give a formal
specification of the structural aspects of the
\emph{WorldClock} micro-architecture using LSL.


\subsection{UML class diagrams}

\begin{figure}[ht]
\vspace{-30\in}
\centerline{\psfig{figure=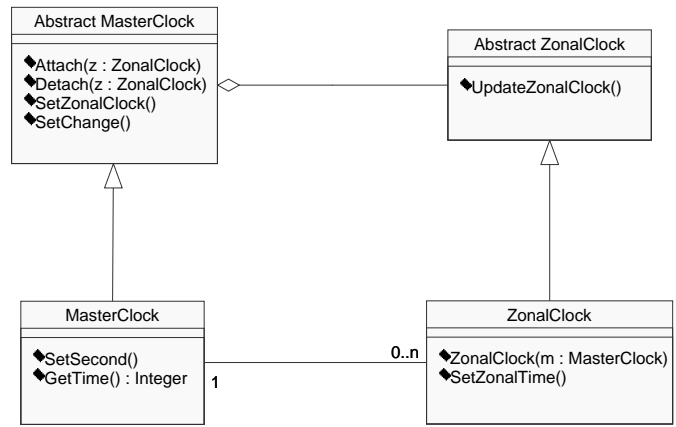,height=2.2in}}
\caption{Participating classes in the \emph{WorldClock} micro-architecture\vspace{-50\in}}
\label{F:cd}
\vspace{-30\in}
\end{figure}

In Fig.~\ref{F:cd}, we show a class diagram of the \emph{WorldClock}
micro-architecture as used for OOD based on UML. This
design could be enhanced to include object references and pseudo-code.

There exists a one-to-many relationship between \emph{MasterClock} and
\emph{ZonalClock} objects. This relationship is expressed by means of
the aggregation arrow in Fig.~\ref{F:cd}. In fact, a
\emph{ZonalClock} object cannot exist independently, and must be
attached to a \emph{MasterClock}, while the converse is not
true. There is a structural aspect which is not expressed in the class
diagram. We require an integrity constraint on the relationship
between the \emph{MasterClock} object and the \emph{ZonalClock}
objects, that is, the \emph{ZonalClock} time must be consistent with
the \emph{MasterClock} time in its zone. This integrity constraint
corresponds to a structural aspect because it characterises the
\emph{states of interest} of the participating objects in the
collaboration.


\subsection{The Larch Shared Language}

In our layered specification approach, LSL serves for the
specification of structural aspects. Let us briefly recall LSL
\cite{GH93} as a specification language. The unit of specification in LSL
is the \emph{trait}. A trait contains a set of operator declarations
(say, a signature), which follows the {\bf introduces} keyword, and a
set of equational axioms, which follows the {\bf asserts}
keyword. Left- and right-hand side of an equation are separated by
``$==$''.  A signature consists of operators the domains and ranges of
which are represented by \emph{sorts}. An equational axiom specifies
the constraints on the defined operators.

The semantics of LSL traits is based on multi-sorted first-order logic
with equality rather than on an initial, final or loose algebra
semantics used by other specification languages
\cite{Bid88,EM85,GT78,ST87}. Each trait denotes a \emph{theory}, i.e.,
a set of logic formulae without free variables, in multi-sorted
first-order logic with equality. The theory contains each of the
trait's equations, the conventional axioms of first-order logic with
equality, everything which follows from them and nothing else.  This
means that the formulae in the theory follow only from the presence of
assertions in the trait---not from their absence.

LSL also provides a way of putting traits together through an {\bf
includes} clause. A trait that includes another trait is textually
expanded to contain all operator declarations and axioms of the
included trait. Boolean operators (true, false, not, $\vee$, $\wedge$,
$\rightarrow$, and $\leftrightarrow$) as well as some overloaded
operators such as if-then-else and ``$=$''\footnote{To avoid
confusion, note that ``$=$'' is the built-in ``equality'' operator
whereas ``$==$'' is used as the connective in equations. By
convention, $t == \mathit{true}$ is written as $t$.  Also note that
``$=$'' binds more tightly than ``$==$''. Otherwise, the two operators
are semantically equivalent.} are built into the language, that is,
traits defining these operators are implicitly included in every
trait.

LSL traits can be augmented with checkable redundancies in order to
verify whether intended consequences actually follow from the axioms
of the trait.  The checkable redundancies are specified in the form of
assertions that are included in the {\bf implies} clause of a trait
and can be verified using Larch Prover~\cite{GH93}. The theory of a
trait can also be strengthened by adding a {\bf generated by} or a
{\bf partitioned by} clause. The {\bf generated by} clause states the
operator symbols that can generate all values of a sort.  The {\bf
partitioned by} clause provides additional equivalences between
terms. It states that two terms are equal if they cannot be
distinguished by any of the functions listed in the clause.


\myfigure{}{%
\parbox{.4\textwidth}{
\begin{lsltrait}
\lsllclindent{0}
\( {\it Time} : \) \key{trait} 
\lsllclindent{1}
\lslskip

\lsllclindent{1}
\key{includes}

\lsllclindent{2}
\( {\it TotalOrder} ( {\it Time} ), \)

\lsllclindent{2}
\( {\it Integer} \) 
\lslskip

\lsllclindent{1}
\( {\it Time} \) \key{tuple} \key{of} 
\lsllclindent{2}
\( {\it hour} , {\it minute} , {\it second} : {\it Int} \) 
\lslskip

\lsllclindent{1}
\key{introduces} 

\lsllclindent{2}
\( {\it currentTime} : \rightarrow {\it Time} \) 

\lsllclindent{2}
\( {\it toInt} : {\it Time} \rightarrow {\it Int} \) 

\lsllclindent{2}
\( {\it fromInt} : {\it Int} \rightarrow {\it Time} \) 

\lsllclindent{2}
\( {\it succ} : {\it Time} \rightarrow {\it Time} \) 

\lsllclindent{2}
\( {\it pred} : {\it Time} \rightarrow {\it Time} \) 

\lsllclindent{2}
\( {\it inc} : {\it Time} , {\it Int} \rightarrow {\it Time} \) 

\lsllclindent{2}
\( {\it dec} : {\it Time} , {\it Int} \rightarrow {\it Time} \) 

\lsllclindent{2}
\( {\it max} : {\it Time} , {\it Time} \rightarrow {\it Time} \) 

\lsllclindent{2}
\( {\it min} : {\it Time} , {\it Time} \rightarrow {\it Time} \) 

\lsllclindent{2}
\( \_\_ \leq \_\_ : {\it Time} , {\it Time} \rightarrow {\it Bool} \) 

\lsllclindent{2}
\( {\it isValid} : {\it Time} \rightarrow {\it Bool} \) 
\lsllclindent{2}
\lslskip
\end{lsltrait}
}\hfill\parbox{.5\textwidth}{
\begin{lsltrait}
\lsllclindent{1}
\key{asserts} 

\lsllclindent{2}
\( {\it Time} \) \key{partitioned} \key{by} \( {\it toInt} \) 

\lsllclindent{2}
${\bf \forall}$ \( t , t_1, t_2 : {\it Time} , h , m , s : {\it Int} , i : {\it Int}
\) 

\lsllclindent{3}
\( {\it isValid} ( {\it currentTime} ) \) 

\lsllclindent{3}
\( {\it isValid} ( t ) == {\it toInt} ( t ) > 0 \) 

\lsllclindent{3}
\( {\it isValid} ( t ) ==  \) 

\lsllclindent{4}
\( ( 0 \leq t . {\it hour} \wedge t . {\it hour} < 24 )\ \wedge \) 

\lsllclindent{4}
\( (0 \leq t . {\it minute}  \wedge t . {\it minute} < 60 )\ \wedge \) 

\lsllclindent{4}
\( ( 0 \leq t . {\it second} \wedge t . {\it second} > 60 ) \) 

\lsllclindent{3}
\( {\it toInt} ( t ) == \)

\lsllclindent{4}
\(\phantom{\mbox{} + \mbox{}} ( 3600 * t . {\it hour} ) \)

\lsllclindent{4}
\(\mbox{} + ( 60 * t . {\it minute} ) \)

\lsllclindent{4}
\(\mbox{} + t . {\it second} \) 

\lsllclindent{3}
\( {\it fromInt} ( {\it toInt} ( t ) ) == t \) 

\lsllclindent{3}
\( {\it succ} ( t ) == {\it fromInt} ( {\it toInt} ( t ) + 1 ) \) 

\lsllclindent{3}
\( {\it pred} ( t ) == {\it fromInt} ( {\it toInt} ( t ) - 1 ) \) 

\lsllclindent{3}
\( {\it inc} ( t , i ) == {\it fromInt} ( {\it toInt} ( t ) + i ) \)

\lsllclindent{3}
\( {\it dec} ( t , i ) == {\it fromInt} ( {\it toInt} ( t ) - i ) \) 

\lsllclindent{3}
\( t \leq t_1 == {\it toInt} ( t ) \leq {\it toInt} ( t_1 ) \) 

\lsllclindent{3}
\( {\it max} ( t_1 , t_2 ) = t_1 == t_2 \leq t_1 \) 

\lsllclindent{3}
\( {\it max} ( t_1 , t_2 ) = t_2 == t_1 \leq t_2 \) 

\lsllclindent{3}
\( {\it min} ( t_1 , t_2 ) = t_1 == t_1 \leq t_2 \) 

\lsllclindent{3}
\( {\it min} ( t_1 , t_2 ) = t_2 == t_2 \leq t_1 \) 
\lslskip

\lsllclindent{1}
\key{implies} 

\lsllclindent{2}
${\bf \forall}$ \( t : {\it Time} \) 

\lsllclindent{3}
\lslskip
\lsllclindent{3}
\( {\it succ} ( {\it pred} ( t ) ) == t \) 
\end{lsltrait}

\vspace{-40\in}
}
}{%
LSL specification of \emph{Time} trait%
}{%
F:Time.lsl%
}

\vspace{-20\in}

\subsection{Abstract states in the \emph{WorldClock} micro-architecture}

We specify the abstract states of \emph{MasterClock} and {\it
ZonalClock} objects using LSL traits. Since both a \emph{MasterClock}
and a \emph{ZonalClock} maintain time, we first specify the \emph{Time}
\emph{sort}.  This is followed by a specification of the \emph{Zone}
sort which provides the data model for \emph{ZonalClock}
objects. Ultimately, we specify the objects in the \emph{WorldClock}
micro-architecture including the relationship between them.

\paragraph{The Time sort}

The LSL specification is shown in Fig.~\ref{F:Time.lsl}. The
\emph{Time} trait includes the traits \emph{TotalOrder(Time)} and
\emph{Integer}. This is shown in the {\bf includes} clause. The
\emph{TotalOrder} trait specifies formally the total ordering of the
abstract values of \emph{Time}. The signature and meaning of the
operator ``$+$'' comes from the \emph{Integer} trait defined in the
LSL trait library. Comparison of time data is defined in terms of
comparison of integers (cf.\ ``$\leq$''). The ``{\bf tuple\ of}''
declaration specifies \emph{Time} data as a record structure. The
signature of the \emph{Time} trait introduces the following functions:
\vspace{-15\in}
\begin{itemize}
\item \emph{currentTime}: Returns the current time.
\item \emph{toInt}: Converts the given time into an integer.
\item \emph{fromInt}: Converts the given integer to time.
\item \emph{succ}: Given a time unit, returns the next time. 
\item \emph{pred}: Given a time unit, returns the previous time.
\item \emph{inc}: Given a time $t$ and an integer $i$,
returns $t$ incremented by $i$ seconds.
\item \emph{dec}: Given a time $t$ and an integer $i$,
returns $t$ decremented by $i$ seconds.
\item \emph{max}: Given two values of time, returns the maximum of the two.
\item \emph{min}: Given two values of time, returns the minimum of the two.
\item $\leq$ : Given two values of time,
tests if the first is not later than the second.
\item \emph{isValid}: Given a value of time,
returns true for a valid time and false otherwise.
\end{itemize}\vspace{-15\in}

\myfigure{}{%
\begin{lsltrait}
\lsllclindent{0}
\( {\it Zone} : \) \key{trait} 
\lslskip

\lsllclindent{1}
\key{includes} \( {\it Integer} , {\it String} , {\it Time} \) 
\lslskip

\lsllclindent{1}
\( {\it Zone} \) \key{tuple} \key{of}

\lsllclindent{2}
\( {\it zonalName} : {\it String}, \) 

\lsllclindent{2}
\( {\it zonalOffset} : {\it Int}, \)

\lsllclindent{2}
\( {\it zonalTime} : {\it Time} \) 
\lsllclindent{2}
\lslskip

\lsllclindent{1}
\key{introduces} 
\lsllclindent{2}

\lsllclindent{2}
\( {\it update} : {\it Time} , {\it Zone} \rightarrow {\it Zone}
\) 

\lsllclindent{2}
\( {\it isUpToDate} : {\it Time} , {\it Zone} \rightarrow {\it Bool}
\) 
\lslskip

\lsllclindent{1}
\key{asserts} 

\lsllclindent{2}
${\bf \forall}$ \( z : {\it Zone} , t : {\it Time} \) 
\lsllclindent{3}

\lsllclindent{3}
\( {\it update} ( t , z ) . {\it zonalName} = z . {\it zonalName} \)

\lsllclindent{3}
\( {\it update} ( t , z ) . {\it zonalOffset} = z . {\it zonalOffset}
\) 

\lsllclindent{3}
\( {\it update} ( t , z ) . {\it zonalTime} = {\it fromInt} ( {\it toInt}
( t ) + z . {\it zonalOffset} ) \) 

\lsllclindent{3}
\( {\it isUpToDate} ( t , z ) == \) \( z . {\it zonalTime} = {\it fromInt}
( {\it toInt} ( t ) \) \( + z . {\it zonalOffset} ) \)
\lslskip

\lsllclindent{1}
\key{implies} 

\lsllclindent{2}
${\bf \forall}$ \( z : {\it Zone} , t : {\it Time} \) 
\lsllclindent{3}

\lsllclindent{3}
\( {\it isUpToDate} ( t , {\it update} ( t , z ) ) \) 
\lsllclindent{3}
\end{lsltrait}
}{%
LSL specification of \emph{Zone} trait%
}{%
F:ZonalClock.lsl%
}

\paragraph{The Zone data model}

The LSL specification of the data model for \emph{ZonalClock} objects
is shown in Fig.~\ref{F:ZonalClock.lsl}. Since a zonal clock should
also maintain a time, the {\em Zone} trait includes the {\em Time}
trait. Besides the zonal time, the {it Zone} sort includes the name
and offset of a zonal clock. Hence, the structure of {\em Zone} is a
{\bf tuple of} three fields. Here is a brief explanation of the
trait's signature:
\vspace{-15\in}
\begin{itemize}
\item \emph{update}: Given a {\it Time} and a \emph{Zone}, returns a
\emph{Zone} whose the zonal-time value is equal to the given {\it Time}
incremented by the offset of the given \emph{Zone}.
\item \emph{isUpToDate}: Given a {\it Time} and a \emph{Zone}, returns
true if the zonal time is up-to-date with respect to the given master
time, and false otherwise.
\end{itemize}\vspace{-15\in}

\myfigure{}{%
\begin{lsltrait}
\lsllclindent{0}
\( {\it WorldClock} : \) \key{trait} 
\lsllclindent{1}
\lslskip

\lsllclindent{1}
\key{includes}

\lsllclindent{2}
\( {\it MutableObj} ( {\it Time} , {\it MasterClock} \) \key{for} \( {\it Obj}
[ {\it Time} ] ) , \)

\lsllclindent{2}
\( {\it MutableObj} ( {\it Zone}, {\it ZonalClock} \) \key{for} \( {\it Obj}
[ {\it Zone} ] ) ) , \) 

\lsllclindent{2}
\( {\it Set} ( {\it ZonalClock} , {\it Set} [ {\it ZonalClock} ] ) \) 
\lsllclindent{2}
\lslskip

\lsllclindent{1}
\key{introduces} 

\lsllclindent{2}
\( {\it masterOf} : {\it ZonalClock} \rightarrow {\it MasterClock} \) 
\lsllclindent{2}

\lsllclindent{2}
\( {\it zonalClocksOf} : {\it MasterClock} \rightarrow {\it Set} [ {\it ZonalClock} ] \) 
\lsllclindent{2}

\lsllclindent{2}
\( {\it isConsistent} : {\it MasterClock} , {\it ZonalClock} , {\it State} \rightarrow {\it Bool} \) 
\lsllclindent{2}
\lslskip

\lsllclindent{1}
\key{asserts} 

\lsllclindent{2}
${\bf \forall}$ \( m : {\it MasterClock} , z : {\it ZonalClock} , {\it st} : {\it State} \) 
\lsllclindent{3}

\lsllclindent{3}
\( \left|{\it zonalClocksOf} ( m )\right| \geq 0 \) 
\lsllclindent{3}

\lsllclindent{3}
\( {\it masterOf} ( z ) = m == z \in {\it zonalClocksOf} ( m ) \) 
\lsllclindent{3}

\lsllclindent{3}
\( {\it isConsistent} ( m , z , {\it st} ) == ( {\it masterOf}
( z ) = m ) \wedge {\it isUpToDate} ( m ! {\it st} , z ! {\it st} ) \) 
\lsllclindent{3}
\end{lsltrait}
}{%
LSL specification of \emph{WorldClock} trait%
}{%
F:WorldClock.lsl%
}

\paragraph{The WorldClock trait}

The LSL specification of the objects in the \emph{WorldClock}
micro-architecture, and the invariant properties of the relationship
between them is shown in Fig.~\ref{F:WorldClock.lsl}. Note that this
is the only trait which deals with (mutable) objects. Both
\emph{ZonalClock} and \emph{MasterClock} objects are defined here.
The other traits {\it Time} and {\it Zone} solely define the data
models. Here is a brief description of the operators declared in the
\emph{WorldClock} trait:
\vspace{-15\in}
\begin{itemize}
\item \emph{masterOf}: Given a \emph{ZonalClock} object, returns the
\emph{MasterClock} object to which it is attached. The
\emph{ZonalClock} object depends on the \emph{MasterClock} object for
its current \emph{zonalTime}. The totality of this function specifies
the constraint that a \emph{ZonalClock} object cannot exist
independently of the \emph{MasterClock}.
\item \emph{zonalClocksOf}: Given a \emph{MasterClock} object, returns
the possibly empty set of attached \emph{ZonalClock} objects that
depend on the \emph{MasterClock}.
\item \emph{isConsistent}: Given a \emph{MasterClock}, a \emph{ZonalClock},
and a {\it State}, returns true if the two clocks are consistently
related to each other in the state, and false otherwise.
\end{itemize}\vspace{-15\in}
\noindent

The specification of the trait relies on the \emph{MutableObj} library
trait that specifies the sort of mutable objects according to the
formal LSL model for objects~\cite{LC95} including a corresponding
notion of \emph{State}. In the axioms, we use the binary operator
``!''  to extract the value of an object from a state. The trait
performs three {\bf includes}. Firstly, we include
``\emph{MutableObj}(\emph{Time}, \emph{MasterClock} {\bf for} {\it
Obj}[\emph{Time}])'' to specify the sort of mutable
\emph{MasterClock} objects whose abstract values are specified by the
\emph{Time} sort. The form ``\emph{MasterClock} {\bf for}
\emph{Obj}[\emph{Time}]'' performs a renaming so that we can use the 
name \emph{MasterClock} for the sort of {\it MasterClock} objects
instead of \emph{Obj}[\emph{Time}]. Secondly, we include
``\emph{MutableObj}(\emph{Zone}, \emph{ZonalClock} {\bf for} {\it
Obj}[\emph{Zone}])'' to specify the sort of mutable \emph{ZonalClock}
objects. Thirdly, we include ``${\it Set} ( {\it ZonalClock} ,
... )$'' so that we can deal with \emph{sets} of \emph{ZonalClock}
objects, namely the set of \emph{ZonalClock}s attached to a
\emph{MasterClock}. The referenced library traits \emph{MutableObj}
and \emph{Set} can be found in~\cite{LC95}.

\vspace{-20\in}

\section{Middle tier: Roles}
\label{S:II}

\vspace{-20\in}

The middle tier in our three-tiered specification approach uses the
data model defined in the lowest tier, and identifies the services
required to specify the roles of the collaborating objects to specify
their externally observable inter-object behaviour. The role
specification for an object includes those services in the interface
of the object which take part and are pertinent to the collaboration
between the object and its collaborators. All the operations of an
object's role are specified using a behavioural interface
specification language (BISL). In the present paper, we have opted for
Larch/C++~\cite{LC95}. While this concrete BISL interacts with C++,
BISLs are also available for other programming languages, e.g., for
Java~\cite{LBR99}.  Hence, conceptually, our approach is
language-independent.


\vspace{-20\in}

\subsection{Informal explanation of services}

Before we discuss the Larch/C++ formalism and we employ it for the
specification of our running example, we explain the services provided
by the \emph{MasterClock} object and the \emph{ZonalClock} objects in
an informal manner. Besides the object which \emph{provides} a
service, we often also need to point out further \emph{involved}
objects. To give an example, the service that creates a
\emph{ZonalClock} object forms part of role specification of the
\emph{ZonalClock} object, and the service involves the
\emph{MasterClock} object to attach the \emph{ZonalClock} object to
the \emph{MasterClock} object. There are the following services:
\vspace{-15\in}
\begin{itemize}
\item \emph{MasterClock}: This object provides an interface for
attaching and detaching \emph{ZonalClock} objects. The services
include the \emph{SetSecond} method to update itself every second, the
\emph{SetZonalClocks} method to send notification to the \emph{ZonalClock}s
when the \emph{MasterClock}'s time changes, the \emph{SetChange} method
to update its time and notify its \emph{ZonalClock}s of time change,
and the \emph{GetTime} method to query the current time corresponding
to the Greenwich Meridian Time.
\medskip
\item \emph{ZonalClock}: This object provides an interface for creation
so that a new instance is attached to a \emph{MasterClock} object, and
it also provides an updating interface to keep its state, namely its
zonal time, consistent with the \emph{MasterClock}'s state via the
methods {\it UpdateZonalClock} and \emph{SetZonalTime}.
\end{itemize}\vspace{-15\in}

Pseudo-code as used in UML would come close to such an informal
definition. In addition, certain UML diagram forms could be used,
e.g., collaboration diagrams or sequence diagrams. In our three-tiered
specification approach, we want to provide \emph{behavioural}
specifications of the services.


\myfigure{!}{%
\begin{lsltrait}

\lsllclindent{0}
{\it MasterClock} : \key{role specification}
\lslskip

\lsllclindent{0}
\key{uses} {\it WorldClock}
\lslskip

\lsllclindent{0}
{\it Attach}({\it z}: {\it ZonalClock})\ \ \{ 
 
\lsllclindent{3}
\key{modifies} $ {\it zonalClocksOf}({\it self});$
 
\lsllclindent{3}
\key{ensures} $z \in {\it zonalClocksOf}({\it self});$

\lsllclindent{1}
\}

\vspace{-30\in}

\lsllclindent{1}

{\it Detach}($z$: {\it ZonalClock})\ \ \{ 
 
\lsllclindent{3}
\key{requires} $z \in {\it zonalClocksOf}({\it self});$
 
\lsllclindent{3}
\key{modifies} ${\it zonalClocksOf}({\it self});$
 
\lsllclindent{3}
\key{ensures} $z \notin {\it zonalClocksOf}({\it self});$

\lsllclindent{1}
\}

\vspace{-30\in}

\lsllclindent{1}

{\it Int} {\it GetTime}()\ \ \{ 
 
\lsllclindent{3}
\key{ensures} $ {\it result} = {\it toInt}({\it currentTime}({\it self}\backslash {\it any}));$

\lsllclindent{1}
\}

\vspace{-30\in}

\lsllclindent{1}

{\it SetSecond}()\ \ \{ 
 
\lsllclindent{3}
\key{modifies} ${\it self};$
 
\lsllclindent{3}
\key{ensures} ${\it self}' = {\it succ}({\it self}\^);$

\lsllclindent{1}
\}

\vspace{-30\in}

\lsllclindent{1}

{\it SetZonalClocks}()\ \ \{ 

\lsllclindent{3}
\key{modifies} ${\it containedObjects}({\it zonalClocksOf}({\it self}), {\it pre});$
 
\lsllclindent{3}
\key{ensures} $\forall z:{\it ZonalClock}\ (z \in {\it zonalClocksOf}({\it self}) \Rightarrow$

\lsllclindent{10}
${\it isConsistent}({\it self}, z, {\it post}));$

\lsllclindent{1}
\}

\vspace{-30\in}

\lsllclindent{1}

{\it SetChange}()\ \ \{ 

\lsllclindent{3}
\key{modifies} ${\it self} \wedge {\it containedObjects}({\it zonalClocksOf}({\it self}), {\it pre});$
 
\lsllclindent{3}
\key{ensures} ${\it self}' = {\it succ}({\it self}\^) \wedge \forall z:{\it ZonalClock}$

\lsllclindent{4}
$(z \in {\it zonalClocksOf}({\it self}) \Rightarrow {\it isConsistent}({\it self}, z, {\it post}));$

\lsllclindent{1}
\}
\lsllclindent{1}
\end{lsltrait}

\vspace{-40\in}
}{%
Role specification of \emph{MasterClock} objects
using the \emph{WorldClock} trait%
}{%
F:MasterClock.lcc%
}

\vspace{-20\in}

\subsection{Behavioural interface specification in Larch/C++}

An object's role specification defines a number of roles (say,
interface functions, or methods for short). To this end, we have to
indicate which trait it \textbf{uses}. This trait provides the names
and meanings of the operators referred to in the definition of the
interface functions. Each such definition consists of a \emph{header}
and a \emph{body}.  The header specifies the name of the interface
function, the names and types of parameters, as well as the return
type (if any). We use the same notation as in
Larch/C++~\cite{LC95}. The body consists of an {\bf ensures} clause as
well as optional {\bf requires} and {\bf modifies} clauses.  The {\bf
requires} and {\bf ensures} clauses specify the pre- and
post-condition respectively. The identifier \emph{self} in the those
assertions denotes the object that receives the corresponding
message. The {\bf modifies} clause lists those objects whose value may
change as a result of executing the method. An omitted {\bf modifies}
clause is interpreted to mean that no object is modified---neither
\emph{self} nor any parameter objects. Finally notice the following
conventions for referring to values and states in role specifications:
\vspace{-15\in}
\begin{itemize}
\item A distinction is made between an object and its value by using an
unannotated identifier (for example, $s$) to denote an object, and a
superscripted object identifier (for example, $s$\^{}\ or\ $s$\'{}) to denote
its value in a state.
\smallskip
\item  The postfix operators ``$\,$\^{}$\,$'' and ``$\,$\'{}$\,$'' for
superscripting are used to extract values from objects. An object
identifier superscripted by ``$\,$\^{}$\,$'' denotes an object's
initial value and an object superscripted by ``$\,$\'{}$\,$'' denotes
its final value.
\smallskip
\item The terms \emph{pre} and \emph{post} refers to the state just before
or after the invocation of the specified method, respectively. The
term \emph{any} can be used when either of these will do. Each of
these has the sort {\it State} which is the sort of the formal model
of states in Larch/C++. If we want to access the value of an object
$o$ in a state $\mathit{st}$, then we use the notation
$o{\setminus}\mathit{st}$, e.g.,
$\mathit{self}{\setminus}\mathit{any}$.
\end{itemize}\vspace{-15\in}

\myfigure{}{%
\begin{lsltrait}

\lsllclindent{0}
{\it ZonalClock} : \key{role specification}
\lslskip

\lsllclindent{0}
\key{uses} {\it WorldClock}
\lslskip

\lsllclindent{0}
{\it ZonalClock}($m$: {\it MasterClock})\ \ \{ 
 
\lsllclindent{3}
\key{contructs} {\it self};
 
\lsllclindent{3}
\key{ensures} ${\it masterOf}({\it self}) = m;$

\lsllclindent{1}
\}

\vspace{-30\in}

\lsllclindent{1}

{\it UpdateZonalClock}()\ \ \{ 
 
\lsllclindent{3}
\key{modifies} ${\it self};$
 
\lsllclindent{3}
\key{ensures} ${\it isConsistent}({\it masterOf}({\it self}), {\it self}, {\it post});$

\lsllclindent{1}
\}

\vspace{-30\in}

\lsllclindent{1}

{\it SetZonalTime}($i$:{\it Int})\ \ \{ 
 
\lsllclindent{3}
\key{modifies} ${\it self};$
 
\lsllclindent{3}
\key{ensures} ${\it self}' = {\it update}({\it fromInt}(i), {\it self}\^);$

\lsllclindent{1}
\}
\lsllclindent{1}

\end{lsltrait}

\vspace{-40\in}
}{%
Role specification of \emph{ZonalClock} objects
using the \emph{WorldClock} traits%
}{%
F:ZonalClock.lcc%
}


\vspace{-20\in}

\subsection{Roles of the clock objects in the \emph{WorldClock}
micro-architecture}

In Fig.~\ref{F:MasterClock.lcc} and Fig.~\ref{F:ZonalClock.lcc}, the
role specifications of \emph{MasterClock} and \emph{ZonalClock}
objects are specified. The formal specifications directly implement
our earlier informal explanations. Note how the operations from the
lowest tier are employed.

\vspace{-20\in}

\section{Highest tier: Interaction}
\label{S:III}

The highest tier employs the lower tiers to provide a specification of
the interaction among the collaborating objects. The interaction
between services provided in their roles must be specified in terms of
operation sequences, and flow of control. These specify the state
transformations of the object collaboration.  In our formal
specification approach, we use a simple designated action calculus for
the interaction layer.

\begin{figure}[ht]
\centerline{\psfig{figure=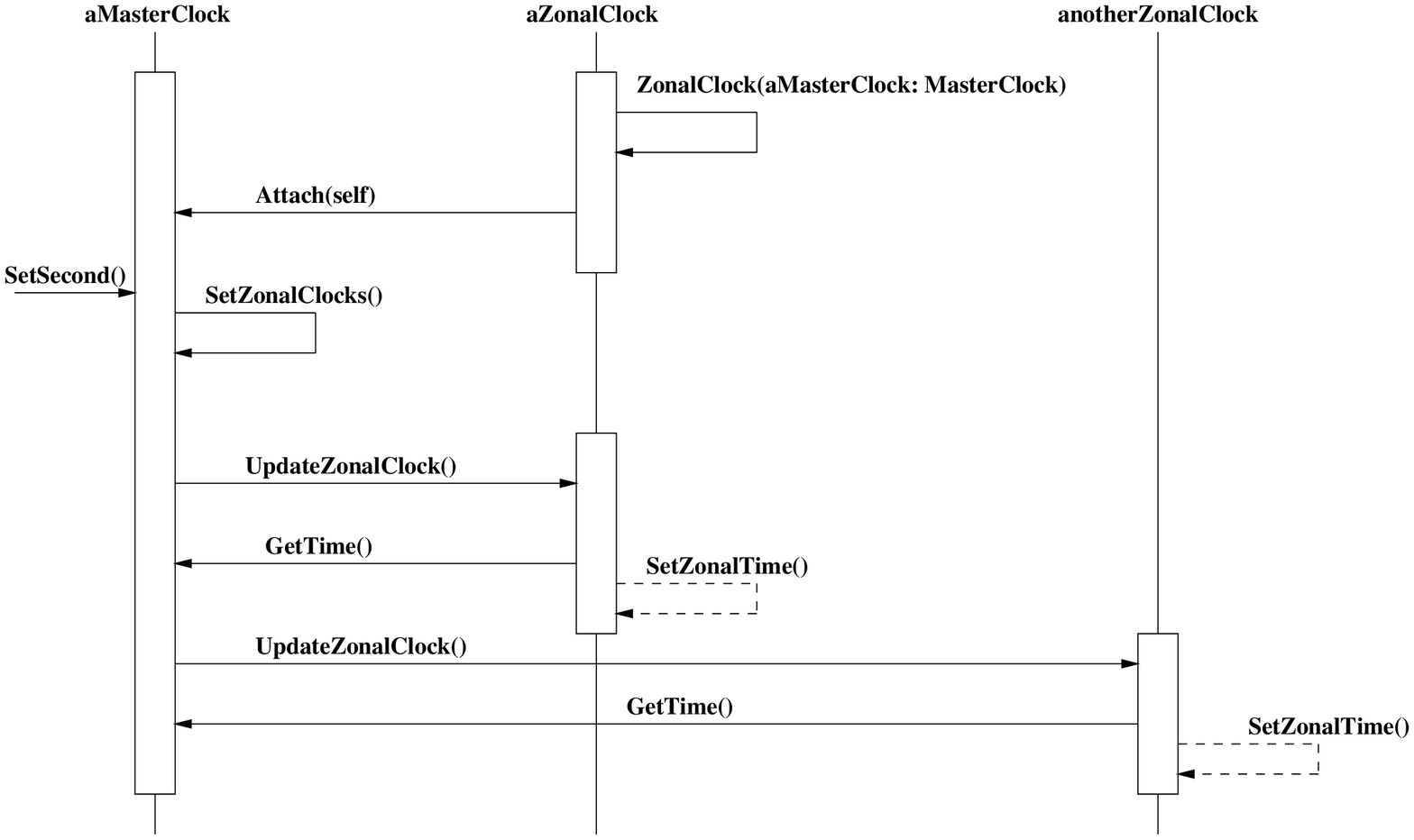,height=2.8in}}
\caption{Sequence diagram for the \emph{WorldClock}
example\vspace{-50\in}}
\label{F:id}
\end{figure}


\vspace{-20\in}

\subsection{Sequence diagrams}

Before we discuss the formal specification of interaction, we recall
the diagrammatic UML approach used in OOD. In Fig.~\ref{F:id}, we
illustrate a collaboration scenario for the \emph{WorldClock}
micro-architecture using a sequence diagram. We illustrate how
\emph{aMasterClock} interacts with \emph{aZonalClock} and
\emph{anotherZonalClock}. It is not possible to capture that there could
be arbitrarily many \emph{ZonalClock} objects, and that the updates
for all \emph{ZonalClock} objects could be done independent of each
other, in parallel. We do not claim that such a diagrammatic is
inherently informal. In fact, there are efforts to assign a formal and
useful semantics to sequence diagrams and other UML notations.
However, an inherent problem with formalising, using, and supporting
UML is its complexity.  We also refer to~\cite{SGJ00} for a critical
review of UML's suitability for the specification of design structures
such as design patterns. We contend that our simple formal approach is
complementary to UML.


\vspace{-20\in}

\subsection{Classification of actions}

The interaction of collaborating objects will be specified using a
logic of actions like Lamport's Temporal Logic of Actions
\cite{Lamport94}. Before we deal with the actual forms of actions, we
want to classify actions to gain a better understanding of
interaction, and of the process in which smaller behaviours contribute
to interactive behaviour.  A \emph{simple} action is in our case an
invocation of a role of an object. Then, a \emph{compound} action is
composed from simpler actions by sequential composition and other
means. Let us categorise actions regarding their possible effects. We
first focus on categories of simple actions, that is, method
invocations, or methods for short. The following categories capture
whether a method affects the state of the associated object, or the
objects in the environment, and whether the execution of the method
returns a value:
\newpage
\vspace{-15\in}
\begin{itemize}
\item A \cv-method returns a \emph{value}.
\item An \co-method \emph{changes} the abstract \emph{state}
of the object itself.
\item An \ce-method \emph{changes} the abstract {\it
state} of the object's \emph{environment}.
\end{itemize}\vspace{-15\in}
The categories \cv\ and \co\ were inspired by a related categorisation
in \cite{BP86}. By an object's environment, we shall mean any object
other than the given object which is related in some way to the given
object.  The differentiated categories \co\ and \ce\ provide a more
precise characterisation of the role behaviour.  We assume that a
method which belongs to the \ce-category must always also belong to the
\co-category, so that any change in the environmental state of an object
is reflected by a change in the abstract state of the object itself.
Further, we assume that methods which both perform state change and
return a value can be reduced to methods which separate these
concerns.  Consequently, we assert that the role specifications from
the middle tier are defined only with actions of the canonical
combinations \co, \co-\ce, and \cv.  To give an example, one can
easily observe the following categories for the role specification of
\emph{MasterClock} objects in Fig.~\ref{F:MasterClock.lcc}:
\vspace{-15\in}
\begin{itemize}
\item \co: \emph{Attach}, \emph{Detach}, \emph{SetSecond}
\item \co-\ce: \emph{SetZonalClocks}, \emph{SetChange}
\item \cv: \emph{GetTime}
\end{itemize}\vspace{-15\in}
Slightly different kinds of categories had to be considered for
compound actions because they are typically concerned with
\emph{several} objects at a time. However, one could still define the 
environment of a given \emph{set} of objects, and one could also
consider the state local to a set of objects. We should assume a
notion of \emph{atomicity} for compound actions, that is, the
intermediate state changes which are caused by steps of execution are
not observable, but only the final state. A value-returning compound
action could also be characterised easily.


\vspace{-20\in}

\subsection{Action combinators}

An interaction specification defines actions on certain collaborating
objects in terms of action combinators like independent or sequential
collaboration.  We group such definitions per \textbf{class}. A single
definition of an interaction is given as a \textbf{method}\
definition.  The object the method of which is invoked and the objects
occurring as method arguments are said to \emph{participate} in an
interaction. There is the following (abstract) syntax for interaction
specifications:

\medskip

{\footnotesize$\begin{array}[t]{lclr}
s & ::= & \textbf{class}\ c\ \{\ i*\ \} & \mbox{(Interaction specifications)}\\
i & ::= & \textbf{method}\ m\ (\ v*\ )\ \{\ a\ \} & \mbox{(Method definitions)}\\
p & ::= & v\ :\ t & \mbox{(Parameters)}\\
a & ::= & & \mbox{(Actions)}\\
  &     & e.m\ (\ e*\ ) & \mbox{(Method invocation)}\\
  &  |  & a \wedge a & \mbox{(Independent composition)}\\
  &  |  & a; a & \mbox{(Sequential composition)}\\
  &  |  & \choice{a}{a} & \mbox{(Composition by choice)}\\
  &  |  & \letin{v}{a}{a} & \mbox{(Substitution)}\\
  &  |  & \ifthen{g}{a} & \mbox{(Guarded action)}\\
  &  |  & \while{g}{a} & \mbox{(Iterated action)}\\
e & ::= & & \mbox{(Expressions)}\\
  &     & v & \mbox{(Instance variable)}\\
  &     & y & \mbox{(Yielder)}
\end{array}\hfill%
\begin{array}[t]{lclr}
c & & & \mbox{(Class names)}\\
m & & & \mbox{(Method names)}\\
v & & & \mbox{(Instance variables)}\\
t & & & \mbox{(Types)}\\
g & & & \mbox{(Guards)}\\
y & & & \mbox{(Yielder)}
\end{array}$}

\medskip

\vspace{-20\in}

This notation immediately suggests that interaction specifications can
be simulated, say executed. A \emph{simple} action is of the form
``$e.m(\ldots)$'', and it denotes the object invocation restricted by
pre- and post-condition of the role specification $m$ for the object
$e$ given in the middle tier. An expression $e$ is either an instance
variable $v$, e.g., \emph{self}, or a \emph{yielder}, that is, the
application of an operator from the lowest tier that yields an object
(reference). We adopt the common convention to omit ``$\emph{self}.$''
in simple actions.

There are three fundamental ways of binary action composition.
\emph{Sequential composition} is used if the order of state changes
matters, either because the actions operate on the same object (cf.\
category \co), or they operate on an overlapping part of the
environment (cf.\ category \ce). By contrast, \emph{independent or
parallel composition} is used if the two actions do deliberately not
interfere each other, that is, they operate on different ``regions''
of the state space. Finally, \emph{choice} selects one out of two
actions depending on which pre-condition evaluates to true. In the
case that both pre-conditions evaluate to true, we deal with
\emph{non-deterministic choice}. The semantics of these forms of
compound actions is defined as follows:
 
\medskip

{\footnotesize$\begin{array}{l|l}
\mbox{Action\ $a_0$}\ &\ \mbox{Pre-/Postcondition} \\ \hline \hline
a_1 \wedge a_2%
\ &\ (\pre(a_0) \Rightarrow \pre(a_1) \wedge \pre(a_2)) \wedge (\post(a_1) \wedge \post(a_2) \Rightarrow \post(a_0))\\ \hline
a_1; a_2%
\ &\ (\pre(a_0) \Rightarrow \pre(a_1)) \wedge (\post(a_1) \Rightarrow \pre(a_2)) \wedge (\post(a_2) \Rightarrow \post(a_0))\\ \hline
\choice{a_1}{a_2}%
\ &\ \phantom{{\vee}}\ ((\pre(a_0) \Rightarrow \pre(a_1)) \Rightarrow (\post(a_1) \Rightarrow \post(a_0)))\\
\ &\ {\vee}\ ((\pre(a_0) \Rightarrow \pre(a_2)) \Rightarrow (\post(a_2) \Rightarrow \post(a_0)))
\end{array}$}%

\medskip

We can generalise the two commutative, associative combinators for
independent composition and choice to \emph{distributed} versions
accepting a \emph{set} of actions as opposed to just two actions.  The
form ``\letin{v}{a_1}{a_2}'' can be regarded as sequential composition
with substitution: every occurrence of $v$ in $a_2$ is substituted by
$a_1$, and then the action $a_2$ is performed. For simplicity, we
might assume that $a_1$ is a simple, purely value-returning action
(i.e., a \cv-method).

There are two more combinators dealing with \emph{guarded} and
\emph{iterated} actions. The action $a_1$ in ``\ifthen{g}{a_1}'' is
executed iff the guard $g$ can be passed. As an aside, parallel
collaboration can be used to define an
``\ifthenelse{g}{\ldots}{\ldots}'' if needed.  The action
``\while{g}{a_1}'' denotes a loop with pre-test. That is, the action
$a_1$ is repeatedly performed as long as the guard $g$ can be
passed. Guards are applications of Boolean operators defined in the
lowest tier. The semantics of {\bf if} and {\bf while} is the
following:

\medskip

{\footnotesize$\begin{array}{l|l}
\mbox{Action\ $a_0$}\ &\ \mbox{Pre-/Postcondition} \\ \hline \hline
\ifthen{g}{a_1}%
\ &\ \phantom{{\vee}}\ ((\pre(a_0) \wedge G \Rightarrow \pre(a_1)) \Rightarrow (\post(a_1) \Rightarrow \post(a_0)))\\
\ &\ {\vee}\ (\pre(a_0) \wedge \neg~G \Rightarrow \post(a_0))\\ \hline
\while{g}{a_1}%
\ &\ \phantom{{\wedge}}\ (\pre(a_0) \wedge G \Rightarrow \pre(a_1))\\
\ &\ {\wedge}\ (\post(a_1) \Rightarrow \pre(a_0))\\
\ &\ {\wedge}\ (\pre(a_0) \wedge \neg~G \Rightarrow \post(a_0))
\end{array}$}

\medskip

This ends our exposition of action combinators for the specification
of interactions between objects. Note how the tiers in our
specification approach are layered. The lowest tier formalises Boolean
operators needed for guards of conditionals and loops in the highest
tier. The lowest tier also provides operators than can be applied to
refer to objects in the interaction specification on the basis of the
structural specification. Furthermore, the middle tier provides the
roles invoked as the simple actions in the highest tier. Finally note
that our set of action combinators for interaction does not include
any form of assignment. State changes are solely encapsulated in the
role specifications.

\myfigure{}{%
\ \ \ $\begin{array}{lcl}
{\bf\ class}\ \mathit{MasterClock}\ \{\\
\ \ {\bf method}\ \mathit{SetZonalClocks}()\ \{\ \bigwedge_{z \in \emph{zonalClocksOf}(\emph{self})}\ z.\emph{UpdateZonalClock}() \}\\
\ \ {\bf method}\ \emph{SetChange}()\ \{\ \emph{SetSecond}(); \emph{SetZonalClocks}()\ \}\\
\\
{\bf class}\ \mathit{ZonalClock}\ \{\\
\ \ {\bf method}\ \mathit{ZonalClock}(m: \emph{MasterClock})\ \{\ m.\emph{Attach}(\emph{self})\ \}
\\
\ \ {\bf method}\ \mathit{UpdateZonalClock}()\ \{%
\begin{array}[t]{l}
{\bf if}\,\neg\ \emph{isConsistent}(\emph{masterOf}(\emph{self}), \emph{self}, \emph{pre})\\
\ \ \ \ \ {\bf then}\\
\ \ \ \ \ \ \ {\bf let}\,i:\emph{Int}\,=\,%
\emph{masterOf}(\emph{self}).\emph{GetTime}()\\
\ \ \ \ \ \ \ {\bf in}\,\emph{SetZonalTime}(i)\ \}
\end{array}
\end{array}$%
}{%
Interaction specification for \emph{WorldClock} example%
}{%
F:interac%
}


\vspace{-20\in}

\subsection{Interaction specification for the
\emph{WorldClock} micro-architecture}

Our running example is completed in Fig.~\ref{F:interac}. In this
particular specification, for every name of an interaction, there
happens to be an interface function of the same name in the role
specifications. Consider, for example, the name \emph{SetChange}.  The
role specification \emph{SetChange} in the middle tier defines the
behaviour in terms of pre- and postconditions whereas the method
\emph{SetChange} in the highest tier defines an \emph{interaction}
based on a sequence of two simpler method invocations. This is a
checkable redundancy where the two specifications focus on different
aspects of the object collaboration. In general, one might define
compound actions which merely \emph{use} services from the role
specifications.

Let us highlight some elements of the specification in order to
illustrate the action calculus. In the specification of
\emph{SetZonalClocks}, the notation ``{\scriptsize{$\bigwedge_{z
\in \emph{zonalClocksOf}(\emph{self})}\ \ldots$}}'' is used for
distributed independent collaboration to point out that the
\emph{UpdateZonalClock} methods can proceed for all the relevant
\emph{ZonalClock}s independently of each other. In the specification of
\emph{SetChange}, sequential collaboration is used because it 
is essential that the \emph{MasterClock} object invokes {\it
SetSecond} and \emph{SetZonalClocks} subsequently on itself. In the
specification of the \emph{UpdateZonalClock}, there is a guarded
action. It models that the role ``$\emph{SetZonalTime}(\ldots)$'' only
needs to be invoked if an update is due. The guard relies on the
operator \emph{isConsistent} from the lowest tier. Also, in order to
retrieve the \emph{MasterClock} object associated to \emph{self}, we
use the operator \emph{masterOf} from the lowest tier. In this manner,
we query a structural aspect of the \emph{WorldClock}
micro-architecture. The guarded action performs ``${\it
SetZonalTime}(i)$'' where $i$ is substituted by the time that is
obtained via an invocation of ``${\it masterOf}({\it self}).{\it
GetTime}()$''.

\vspace{-20\in}

\section{Related work}
\label{S:related}

\vspace{-20\in}

A good specification of the micro-architectures in a framework is
indispensable to both the reuser of the framework and to the
developer/maintainer of the framework. Yes, it is like saying that
``every program must have a specification''. If a reuser of the
framework has to understand the framework, either the reuser could
look at its underlying design patterns (which are very abstract), or
(s)he could read the code to extract some behavioural insight (which
is a waste of time). We contend that \emph{specifications} of
micro-architectures \emph{complement} high-level designs such as UML
diagrams, and that this marriage explains the behaviour of objects in
the framework in an appropriate manner. Let us consider previous work
on specification of object-oriented designs.

In~\cite{LK98}, UML notation for class and sequence diagrams is
enriched by certain precise visual constraints which are useful for
design-pattern descriptions. In order to specify, for example, that a
certain participating class can occur several times in an actual
pattern instance, suitable Venn diagrams are used.  This is an
improvement over the style used in the GoF catalogue~\cite{GHJV94}
where such constraints are treated in an informal manner, e.g., by
giving an example with two sample classes. The enriched UML sequence
diagrams covers in part the aspects captured in the middle and the
highest tier in our approach. However, behavioural specifications
are not an issue, neither does this UML approach adhere to a layered
specification discipline.

In~\cite{LBG96}, various design patterns are formally proved to be
refinement transformations in a semantical sense when compared to a
more native/hard-wired encoding of the corresponding design
pattern. To this end, an object calculus theory~\cite{FM91} is used as
the semantical framework. This formal model does not aid the reusable
behavioural specification of micro-architectures, nor does it address
the issue of object interaction. The work is geared towards a
representation of design patterns in formal specifications for
purposes of semantical reasoning.

In~\cite{KKS96}, a logic on parse trees is used to specify and enforce
constraints on a design language. One can deal with architectural
characteristics, or even with source code requirements in this
manner. One possible application is to use the logic in order to
enforce invariants of design-pattern instantiations. One can also use
the mainstream language OCL to describe certain well-formedness
constraints on an object-oriented design or a program. Such approaches
emphasise structural or even syntactical invariants of
micro-architectures, design patterns, or styles. Specification of
behaviour and interaction is beyond the scope of these approaches.

In~\cite{Mikkonen98}, design patterns are formally specified using the
formal specification method DisCo for reactive systems. The formal
basis of DisCo is Temporal Logic of Actions~\cite{Lamport94}. In this
setting, actions are understood as multi-object methods. These actions
are atomic units of execution. An action consists of a list of
participants and parameters, an enabling condition, and the definition
of state changes caused by an execution of the action. An important
contribution of this work is that it indicates how combination and
instantiation of patterns in terms of their defining multi-object
methods can be performed. The DisCo specifications roughly correspond
to the interaction specifications in the highest tier. However, in our
approach, it is essential that the interaction specifications are
defined on top of the other layers for structural and behavioural
aspects of micro-architectures.

In~\cite{Eden00}, a specification approach is described to capture
design patterns and other building blocks. To this end, a declarative
specification language LePUS corresponding to a subset of higher-order
logic is employed. The prime idea is to define (say, constrain)
object-oriented designs and their building blocks in terms of suitable
sets of methods and classes, and relations on these. A non-trivial
example is a so-called tribe which is a set of clans, that is, a set
of methods which share a signature, in relation to a set of
classes. Such a tribe is relevant in the specification of the
\emph{Visitor} pattern~\cite{GHJV94}. This approach addresses
non-trivial structural properties but it is not suited to specify the
behaviour of collections of objects, neither does it address object
interaction.

In~\cite{FMR01}, object-oriented designs and design patterns are
specified using RSL---the RAISE specification language~\cite{RAISE}.
This approach addresses weaknesses of informal and diagrammatic
approaches in that it helps a designer to demonstrate conclusively
that a particular problem matches a particular pattern, or that a
proposed solution is consistent with a particular pattern. The
proposed type of specification clearly captures more structural
properties than a pure class diagram. The formal model also attempts
to capture some behavioural characteristics such as the variables
changed by a certain method invocation. Still the approach is rather
syntactical in that it merely defines well-formedness relations on
representations of object-oriented designs. Also, the approach
neglects object interaction.

In~\cite{HHG90}, a significant contribution to the problem of
specifying micro-architectures is presented. The authors present a
modelling construct called \emph{contracts} for the specification of
behavioural compositions. The paper illustrates that behavioural
specifications are meaningful for refinement, conformance testing, and
instantiation. The specification language is structured in the sense
that different kinds of obligations or constraints are considered,
namely typing, behavioural and contractual obligations. When compared
to our specification language, contracts do not adhere to the
multi-layered principle. Also, no formal syntax and semantics has been
given for the specification language. This hampers tool support, and
the verification of the correctness of programs that implement the
micro-architectures.

To summarise, previous approaches usually focus on only selected
aspects of collections of collaborating objects while our approach
identifies three different tiers to achieve full coverage. Most
previous approaches involve informal ingredients. This is particularly
true for the mainstream UML approach. By contrast, our simple formal
approach is accessible for formal testing and verification. That is,
Larch/C++ specifications can be tested~\cite{Celer95}, and properties
defined in terms of LSL traits can be verified~\cite{GH93}. Although
our specifications are formal, they are immediately useful in the
programming phase as supported by the Larch tool suite. Previous
approaches usually emphasise the specification of design patterns.
Note that \emph{behavioural} specifications are not too much of an
issue for design patterns. Even ``behavioural'' patterns~\cite{GHJV94}
are rather ``algorithm-free'': their vague operational meaning is
usually indicated via pseudo-code, and it can hardly be captured with
specifications using pre- and postconditions or otherwise. By
contrast, we focus on micro-architectures. These building blocks of
domain-specific application frameworks usually exhibit some
interesting behaviour subject to role specifications in our middle
tier.

\vspace{-20\in}

\section{Conclusion}
\label{S:conclusion}

\vspace{-30\in}

\paragraph{Three-tiered specification}

Our approach enables the reuser to understand the structural aspects
(lowest tier), the behavioural aspects (middle tier) and the
interactive aspects (highest tier) of collections of collaborating
objects. The specification of behaviour in terms of roles depends on
the specification of structural aspects. That is, the operators from
the lowest tier are used in the pre- and postconditions in the middle
tier.  The specification of interaction between the collaborating
objects in the highest tier invokes the roles of the objects specified
in the middle tier, and operators from the lowest tier are used to
query structural aspects. The three tiers achieve a separation of
concerns. In each tier, a designated notation is favoured. Structural
aspects are preferably specified in algebraic style as supported by
LSL. Behavioural aspects are best described by contracts based on pre-
and postconditions as supported by Larch/C++. Finally, the interaction
of collaborating objects is best captured by an action calculus, say,
a TLA-like logic. The approach adheres to a layered architecture
principle.  Lower tier specifications are imported into the next
higher tier. Inclusion of a component from a higher tier to a
component in a lower tier is not permitted, that is, there are no
up-calls. This design permits improved reuse of components, and it
allows changes to components in one layer without affecting the
components in lower layers.  In previous
studies~\cite{AAM01,AZ01,AHO02}, we have investigated the virtues of
different variations on a three-tiered system architecture, and we
have shown its expressiveness for the development of real-time
reactive systems, E-Commerce systems, and evolving systems.  For
instance, the three tiers of a reactive system architecture
respectively contain specifications of abstract data types, reactive
classes, and system configurations. The \emph{contribution} of the
present paper is to capture our experience in a three-tiered
specification approach.

\vspace{-30\in}

\paragraph{Evolving systems}

Let us indicate how our specification approach enables the evolution
of a software system when requirements are revised or new ones need to
be added. This will further clarify the usefulness of the
specification approach in practice. System evolution is partitioned
horizontally as well as vertically: evolution can happen in each tier,
and propagation of change is across two successive tiers. To slightly
generalise our \emph{WorldClock} example from before, let us consider
a system which involves publishers and subscribers instead of
\emph{MasterClock}s and \emph{ZonalClock}s. This more abstract scenario
corresponds to the \emph{Observer} pattern~\cite{GHJV94}, also known
as {\em Publisher-Subscriber} pattern. The pattern basically suggests
that a publisher notifies any number of subscribers about changes to
its state.  Our layered specification approach offers the following
advantages if a system involving publishers and subscribers evolves:
\vspace{-15\in}
\begin{itemize}
\item Traits can be refined for whatever reason of evolution in the
first tier, and a refined trait may be included in a role
specification to enrich the theory or strengthen data structuring.  To
give a simple example, if it is required for a publisher to maintain
an ordered list of currently subscribed components, the {\em Set}
trait can be refined to {\em List} in the first tier, and the {\em
List} trait is then linked to the role specification. In general,
changes which are local to the first tier, can be analysed prior to
their propagation to the second tier. Refinement mappings for Larch
traits are discussed in \cite{Colagrosso93}.
\item An object can take both roles, that of a publisher as well as a
subscriber, because in principle an object can subscribe to several
publishers, and an object subscribing to a publisher may itself be a
publisher for some other objects. If such a combination of
responsibilities or services is required, then this is easy to achieve
because role specifications in the second tier can be linked to
several traits via the \emph{uses} clause. Without layering, such
requirements are difficult to model in a formal manner.
\item The publisher may decide which internal state changes it wants
to share with its subscribers. It may also decide when and how to
communicate such changes.  These decisions are manifested in the role
specification of the publisher. A revision of these decisions only
requires local modifications in the middle tier. The interaction
specification in the third tier will be ``automatically'' updated with
the modified methods. In~\cite{AAM01}, a theory is given for
constructing composite classes and class refinements. Evolution in the
second tier would need to satisfy a similar theory.
\item Suppose the requirement of the publisher evolves such that the
publisher is required to communicate state changes to its subscribers
without the subscribers knowing the identity of the publisher. Then,
we need to introduce a new role specification in the second tier, say
{\em Channel}. In addition, the interaction components in the third
tier have to be revised as follows:
\begin{itemize}
\item the publisher interacts with the channel, and
\item the channel interacts with subscribers.
\end{itemize}
\end{itemize}\vspace{-15\in}

\vspace{-30\in}

\paragraph{Towards an integrated development method}

Each tier in our specification approach gives rise to an
implementation layer. Also, the presented specification approach
allows for a seamless integration of the three tiers with UML visual
modelling facilities, automated testing of Larch/C++
specifications~\cite{Celer95}, and verification of properties defined
in terms of LSL traits~\cite{GH93}. The components in each layer can
be individually analysed before before composition, adaptation, or
reuse. It turns out that one important notion is missing in our
layered approach to the specification of micro-architectures: we lack
a sufficiently expressive and automated model for reuse of class
structures. That is, to actually deploy reusable micro-architectures
or even design patterns in an actual application context, we need
language constructs, tool support, and other means to adapt library
structures, to replicate participants in a collaboration, to refine
micro-architectures, and to compose behaviours of
micro-architectures. In our ongoing research, we attempt to reuse
ideas from programming support for design
patterns~\cite{BHK98,LSDF97,Bosch97,Bosch98,BFLS99,FL00} and
corresponding ideas for tool support for
OOA/OOD/OOP~\cite{FMW97,SGJ00,FLM01}.  In particular, we plan to base
the syntactical notion of reuse on \emph{superimposition of class
structures} as defined in our previous work~\cite{BFLS99,FL00,FLM01}.
This will eventually lead to a fully integrated software development
method centered around the notion of micro-architectures.


\vspace{-20\in}

\paragraph{Acknowledgement}

The first author is grateful for the collaboration with Sridhar
Naraya\-nan who contributed to the subject of the paper in an
important way.

\vspace{-30\in}
 
{\small
\bibliographystyle{abbrv}
\bibliography{report}
}

\end{document}